\documentclass[10pt,two column]{article} 

\usepackage[a4paper,bindingoffset=0.2in,left=0.5in,right=.5in,top=.5in,bottom=1in,footskip=.5in]{geometry}
\usepackage{blindtext}

\usepackage{times}
\usepackage{graphicx}
\usepackage{amssymb}
\usepackage{url,hyperref}
\usepackage{epsfig}
\usepackage[T1]{fontenc}
\usepackage{amssymb}
\usepackage{amsmath}
\usepackage{amsfonts}
\usepackage{verbatim}
\usepackage{graphicx}
\usepackage{hyperref}
\usepackage{algorithmic}
\usepackage{algorithm}
\usepackage{array}
\usepackage{multirow}
\usepackage{epstopdf}
\usepackage{color}
\begin{document}

\title{FPGA based High Speed Data Acquisition System for High Energy Physics Application}

\author{Swagata Mandal, Suman Sau, Amlan Chakrabarti, Subhasis Chattopadhyay\\
\\
VLSID-2015, Design Contest track, Honorable Mention\\
January 3-7, 2015 \\
\\
University of Calcutta, VECC Kolkata\\
Kolkata, India\\
\\
sumansau@gmail.com\\
}

\date{}
\maketitle

\begin{abstract}
In high energy physics experiments (HEP), high speed and fault resilient data communication is needed between detectors/sensors and the host PC. Transient faults can occur in the communication hardware due to various external effects like presence of charged particles, noise in the environment or radiation effects in HEP experiments and that leads to single/multiple bit error. In order to keep the communication system functional in such a radiation environment where direct intervention of human is not possible, a high speed data acquisition (DAQ) architecture is necessary which supports error recovery.
\par
This design presents an efficient implementation of field programmable gate array (FPGA) based high speed DAQ system with optical communication link supported by multi-bit error correcting model. The design has been implemented on Xilinx Kintex-7 board and is tested for board to board communication as well as for PC communication using PCI (Peripheral Component Interconnect express). Data communication speed up to 4.8 Gbps has been achieved in board to board and board to PC communication and estimation of resource utilization and critical path delay are also measured.   
\end{abstract}
\section{Motivation}
There remains an immense challenge in developing an efficient DAQ chain for HEP experiments as there is a demand of high data rate, low error and scope for further development of system architecture. The DAQ chain in general consists of analog sensor hardware followed by analog to digital (A/D) converter with high resolution, that gets connected to digital part of the DAQ chain for storage and further processing. In our work, we have targeted the digital part of the DAQ chain, which communicates with the host computer for further analysis of data at the back end.
In general a successful HEP experiment requires a DAQ chain to handle the following issues:
\begin{itemize}
\item \# channels >100k
\item Read-out  frequency > 100 kHz
\item Synchronization limit < 100 ps
\item Data Capacity > 1 Tb/s
\end{itemize}
We have considered FPGA for our hardware prototype development due to its reconfigurability, which perfectly supports an evolving design requirement, as well as due to availability of design IP and flexibility of protocol implementation in terms of hardware software co-design. Our proposed system provides high data rate with transient error correction capability. 

\section{Design Requirement} For the DAQ prototype design described in this paper we have used two Xilinx Kintex 7 (KC705) boards, one optical fiber cable, jitter cleaned clock generator (CDCE62005EVM), power supply (220V), one host PC, Xilinx ISE 14.5 software with Chipscope Pro Analyzer tools.

\section{System Design } \label{FunctionalBlockOfSystem}
The complete flow of the system with different functional blocks are shown in Figure~\ref{fig:BlockDiagramFlow}. A detail functional description of each of the block and their importance are described in this section. In our prototype we have taken 48 bit input along with 4 bit slow control field as data, which is transmitted over the optical link.
\subsection{Scrambler/De-scrambler} Scrambler is used here to reduce the occurrence of long sequences of `1' (or `0') that maintains a good DC balance in input signal coming from the A/D converter. It is used to enable accurate timing recovery on receiver equipment without resorting to redundant line coding. It has a latency of one clock cycle but does not add any redundancy in the system unlike the $8b/10b$ or $7b/8b$ line coding. De-scrambler just does the opposite with respect to the scrambler in the receiver side.
\subsection{BCH Encoder/Decoder} BCH is a binary error correcting code~\cite{Bose196068}. This functional block is used to detect the errors, which occur due to the radiation (SEU~\cite{Nashiyama:SEU}) or hits by the charged particles during the transmission.
Here, we have used (15,7,2) BCH code which can correct up to two bits error. In this coding scheme 7 bit data is appended with 8 redundant bits for error correction. So the coding efficiency is 0.467. To increase the coding efficiency, we can use (15,11,1) BCH code but in that case error correcting capability will be reduced from two to one. Similarly, we can use triple error correcting BCH code~\cite{Xie:Tripple:BCH} but that will reduce coding efficiency. To optimize between coding efficiency and error correcting capability we have chosen (15,7,2) BCH code. This block increases an extra hardware redundancy and clock latency in the whole system that increases reliability in the data transfer. For decoding of the coded data we follow three steps:
\begin{enumerate}
    \item Determination of the error locater polynomial
      \item Detection of error location using Chien Search Algorithm~\cite{chien:search}
            \item Location of the data at the error position
      \end{enumerate} 
\begin{figure*}[htb]	
\centering
\includegraphics[scale=0.35]{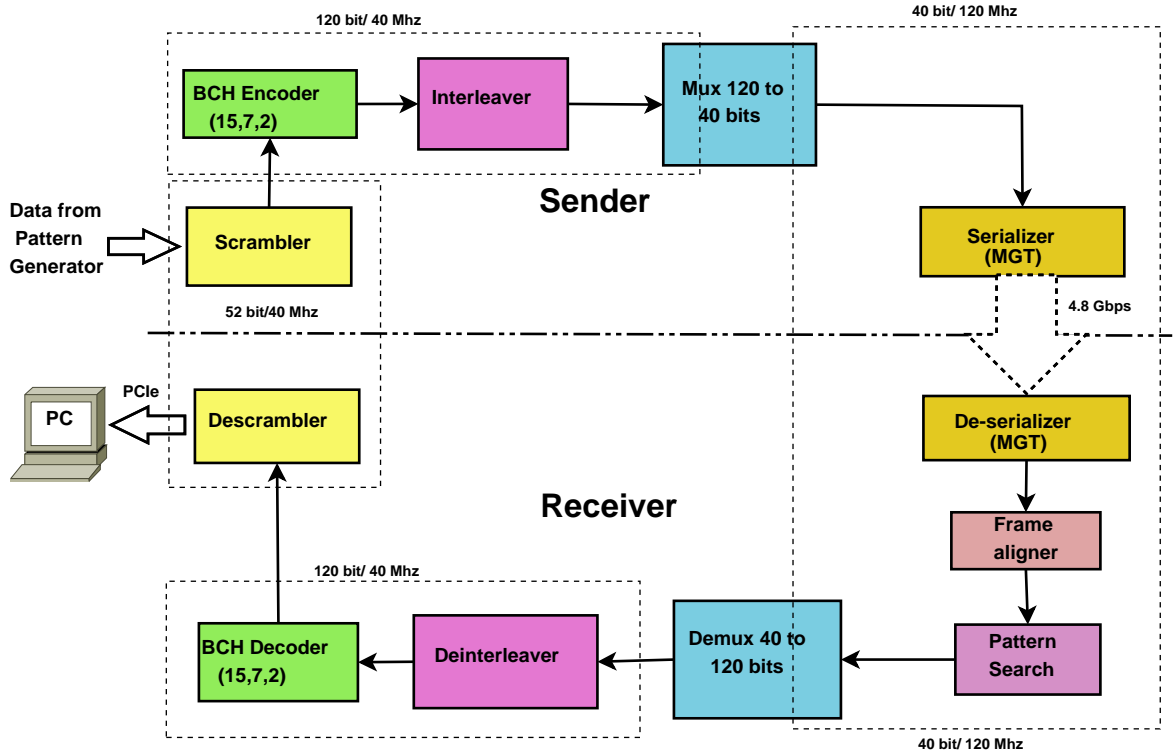}
\caption{Internal blocks of the proposed system}
\vspace{-8pt}
\label{fig:BlockDiagramFlow}
\end{figure*}
\subsection{Interleaver/De-interleaver}\label{Interleaver} Interleaving is the reordering of the data that is to be transmitted, so that the consecutive bytes of data are distributed over a larger sequence of data to reduce the effect of burst error. The use of interleaving greatly improves the capacity of the code to correct burst errors. Normally two kinds of interleaving are used in any communication system:
\begin{enumerate}
    \item Block interleaver
      \item Convolutional interleaver
\end{enumerate} 
 Here we have used the block interleaver. This interleaver block does not add any extra clock latency in the whole system. The whole process increases the code correction capabilities without any extra overhead. 
\begin{figure}[htb]
\centering
\includegraphics[scale=0.35]{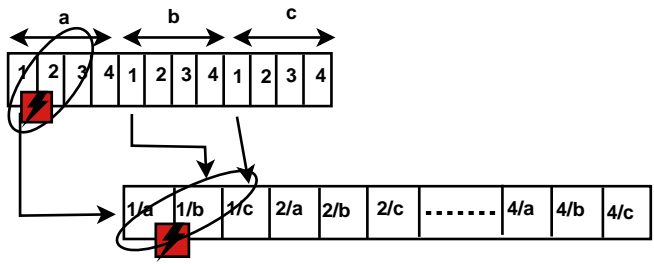}
\caption{Interleaving process and effects of burst error}
\vspace*{-8pt}
\label{fig:Interleaver}
\end{figure}
 Figure~\ref{fig:Interleaver} shows interleaving process taking three blocks of data (each block size is 4 bit) into consideration. During the transmission, if any noise disrupts 4 bits of data, the errors are distributed in the received data. So there is a marginal amount of distortion instead of completely loosing the data in the received block during the burst errors. De-interleaver process just does the opposite with respect to the interleaver in the receiver side.
\subsection{MUX/DEMUX and Clock Domain Crossing} 
\begin{figure}[h]
\centering
\includegraphics[scale=0.25]{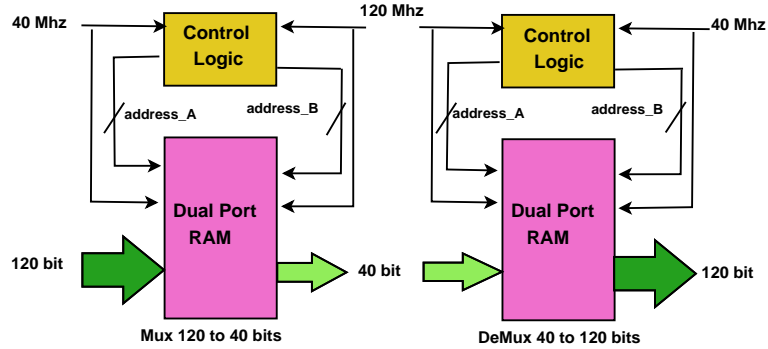}
\caption{Mux DeMux for clock domain crossing}
\vspace*{-8pt}
\label{fig:MuxDMux}
\end{figure}

This block consists of dual port RAM and read write controller. It breaks down 120 bit frame into three words of 40 bits width. Here, we have used 120 MHz clock to drive the multi-gigabit transceiver (MGT). The data rate and clock frequency can be changed to any value according to the requirement. This block is used to synchronize the data rate between MGT and the other parts of the design keeping the data rate same. It also reduces the bandwidth consumption in the channel. It is used both in transmitter and receiver side for synchronization. Figure~\ref{fig:MuxDMux} shows the architectural block diagram of the MUX-DEMUX and clock domain crossing.
\subsection{Serializer/De-serializer} This block simply converts the parallel data to serial data, which is transmitted over the communication channel. It is inbuilt within the MGT.  De-serializer simply converts the serial data to parallel data in the receiver side.
\subsection{Frame Aligner and Pattern Search} \label{FrameAligner}
\begin{figure}[htb]
\hspace{-20 pt}
\includegraphics[scale=0.27, angle=0]{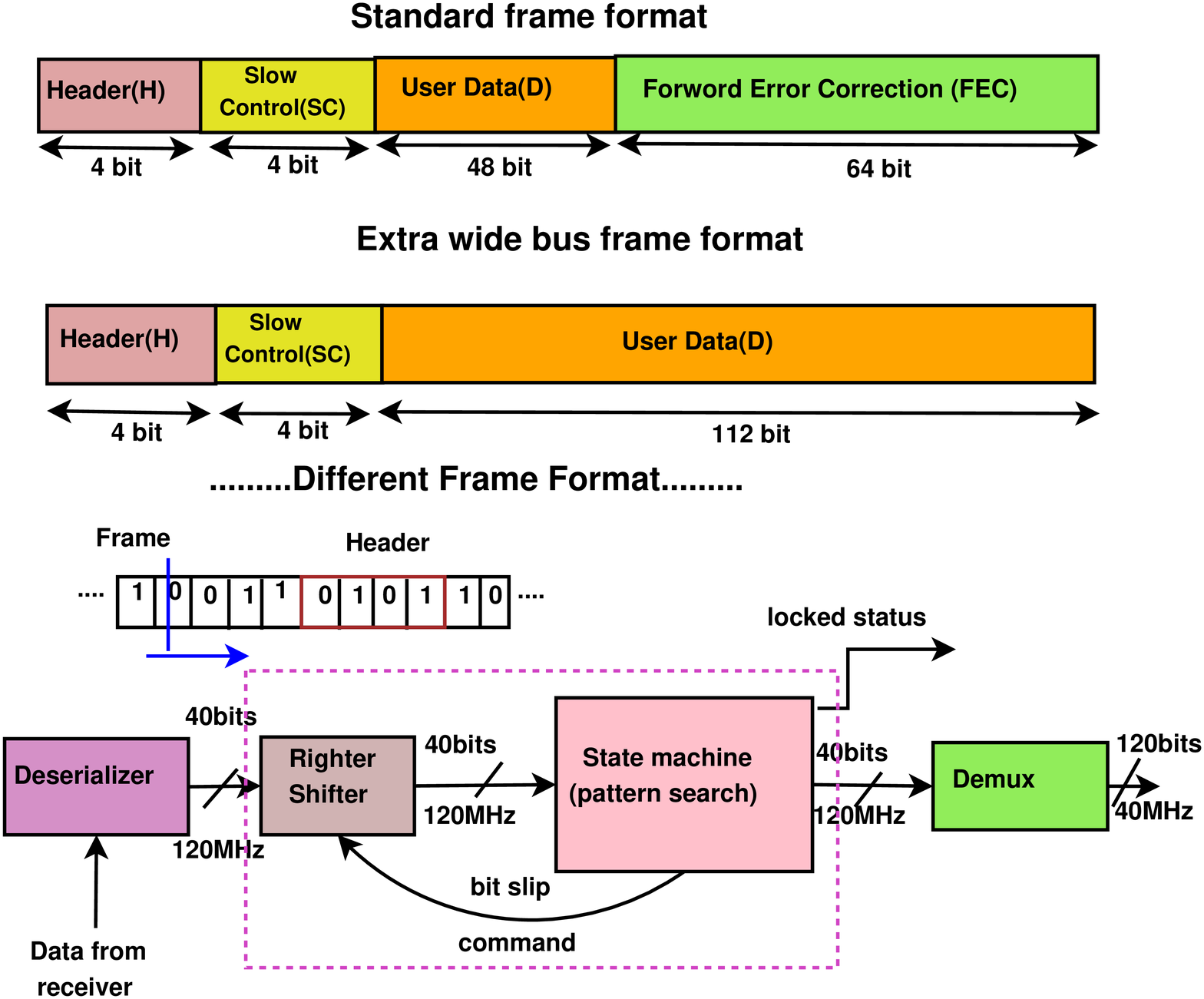}
\caption{Different frame formats and Frame Alignment process}
\vspace*{-8pt}
\label{fig:FrameAlignerWork}
\end{figure}

The frame aligner block is only used in the receiver side. Data may be affected by noise when transmitted through the channel. Hence, the frame aligner aligns the frames correctly before further processing. Every frame has a frame header, which is used to detect a frame type properly and is to be searched first in the receiver side. Different frame formats of our system are shown in top of the Figure~\ref{fig:FrameAlignerWork}. The standard frame consists of four fields: Header field ($H$) consist of 4 bit data, Slow Control ($SC$) field consists of 4 bit data, Data ($D$) field of width 48 bit, Forward Error Correction field of width 64 bit ($FEC$). 
$SC$ field is reserved for controlling the DAQ chain in future.
  Whereas in extra wide bus frame first two fields are same but there is no such error correction and width of the data field is 112 bit. So extra wide bus frame consists of three field: Header field, Slow control field and data field.  Extra wide bus frame format will be used for those applications where probability of occurrence of error is very less like out side the radiation zone. Thus the efficiency of data transfer is higher in extra wide bus frame format compromising with the errors.
The frame aligner and pattern search block consists of two sub blocks (Pattern search block, Right shifter block) as shown in bottom of the Figure~\ref{fig:FrameAlignerWork}. Right shifter block shifts the receive data by one bit to the right side from MSB side and send it to pattern search block. The pattern search block checks whether the header is received or not. Once the header is properly detected pattern search block will continuously search for header for another 32 times and then the search process will be completed and frame becomes synchronized. Until the header is properly detected a bit slip command will be generated and the searching process for the header will be going on.   


\subsection{Data Transfer to Host PC through PCIe}
\begin{figure*}[htb]
\centering
\includegraphics[scale=0.22]{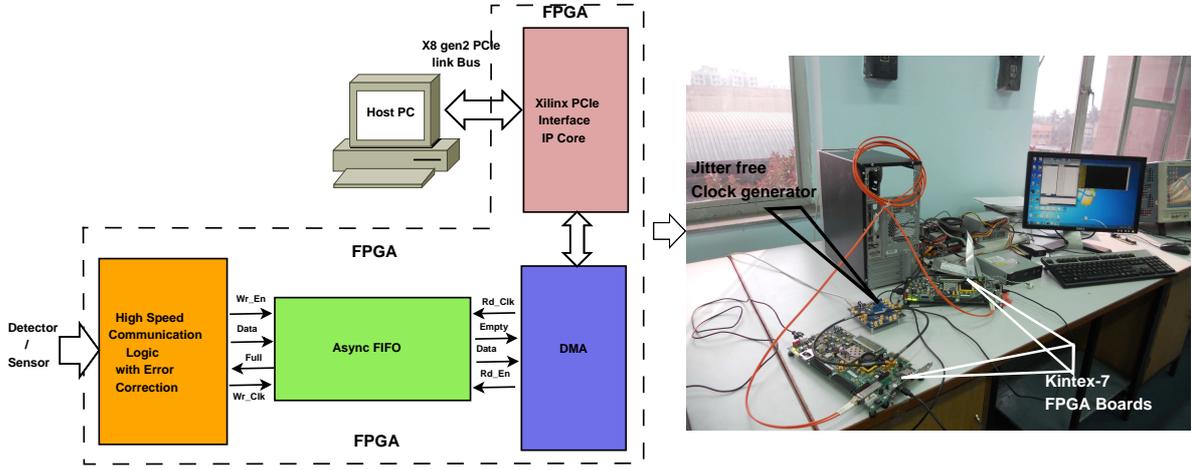}
\caption{ PCIe interfacing with blocks and Experimental setup of proposed DAQ}
\label{fig:PCIeSetup}
\end{figure*}

An asynchronous Fast In Fast Out (FIFO) and Scatter Gather Direct Memory Access (SGDMA) is used to transfer data from FPGA board to PC through PCIe. We have used PCIe gen 2 IP core available from Xilinx. Interconnection of FPGA to PC through PCIe is shown in Figure~\ref{fig:PCIeSetup}. Data will be written in to FIFO at a frequency (120 MHz) by which MGT is running and  data will be read from FIFO at a frequency (125 MHz) by which PCIe core is running. In the PC side we capture the data by a program had been developed using windows software development kit (SDK) written in C language.  

\par
\begin{figure*}[htb]
\centering
\includegraphics[scale=0.27]{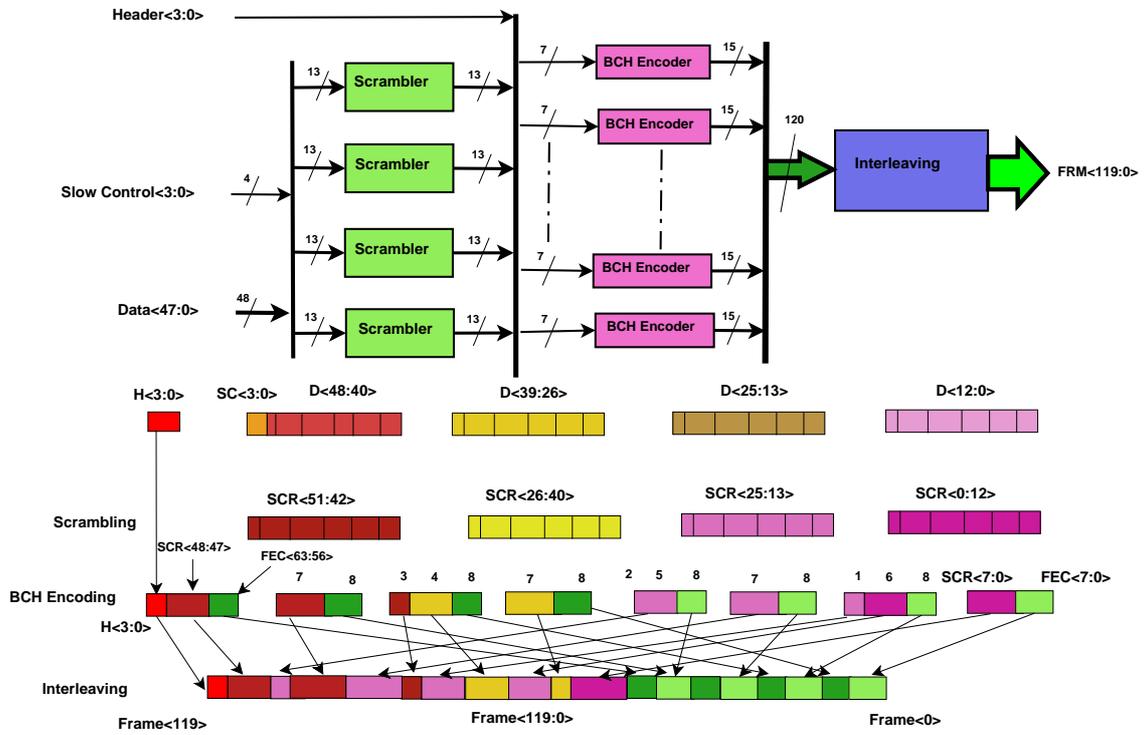}
\caption{Standard frame format generation}
\label{fig:data_flow}
\end{figure*}

 The complete chain of the functional blocks as shown in Figure~\ref{fig:BlockDiagramFlow} for the high speed DAQ with multi-bit error correction (here we take up to two bits error correction) has been implemented on the FPGA board. Figure~\ref{fig:data_flow} shows the complete serial flow diagram of the generation of a standard frame format. At first, only 52 bit user data is scrambled by the scrambler block. This 52 bits data is divided into four 13 bits block of data and scrambles each block parallely. The scrambled data with the 4 bit header is mapped in the input lines of the eight BCH (15,7,2) encoder. Here, each BCH encoder block can correct up to 2 bits of error within 7 bits of input. So the total  $8\times 2 = 16$  bits can be corrected using this technique with out any extra resources. Output of all the encoders are appended to get a frame of 120 bits data. This 120 bits of data is interleaved first and then goes to the next functional block that is the MUX. Interleaving (described in section~\ref{Interleaver}) is used to reduce the effect of burst error. But the header ($H$) position which would not changed in the frame format (red color in Figure~\ref{fig:data_flow}) even after interleaving process, helps to synchronize the frame in the receiver side. In Mux-DeMux and clock domain crossing block a dual port RAM is used to write this 120 bits data using 40 MHz clock and read the same data in 120 MHz clock rate and 40 bit word size. So the data rate  for writing is $40 \times 120= 4.8 $ Gbps  and for reading is $120\times 40=4.8$ Gbps that are same. The 40 bit data is serialized first and goes to the transmitter (TX) for transmitting over the optical fiber cable. In the receiver (RX) side functional blocks are De-serializer, DEMUX and clock domain crossing, De-interleaver, BCH Decoder (15, 7, 2) and De-scrambler. They performs just opposite function with respect to Serializer, MUX, Interleaver, BCH Encoder (15, 7, 2) and Scrambler. Only the Frame Aligner and Pattern Search blocks are the extra component in the receiver side. The detailed functional description of these  extra blocks are given in  the section~\ref{FrameAligner}. For standard frame format generation, we have used $1010$ as header and for the extra wide bus frame $0101$ is used as header. 
\section{System Implementation} \label{ExperimentalVerification}
\begin{figure*}[htb]
\centering
\includegraphics[scale=0.70]{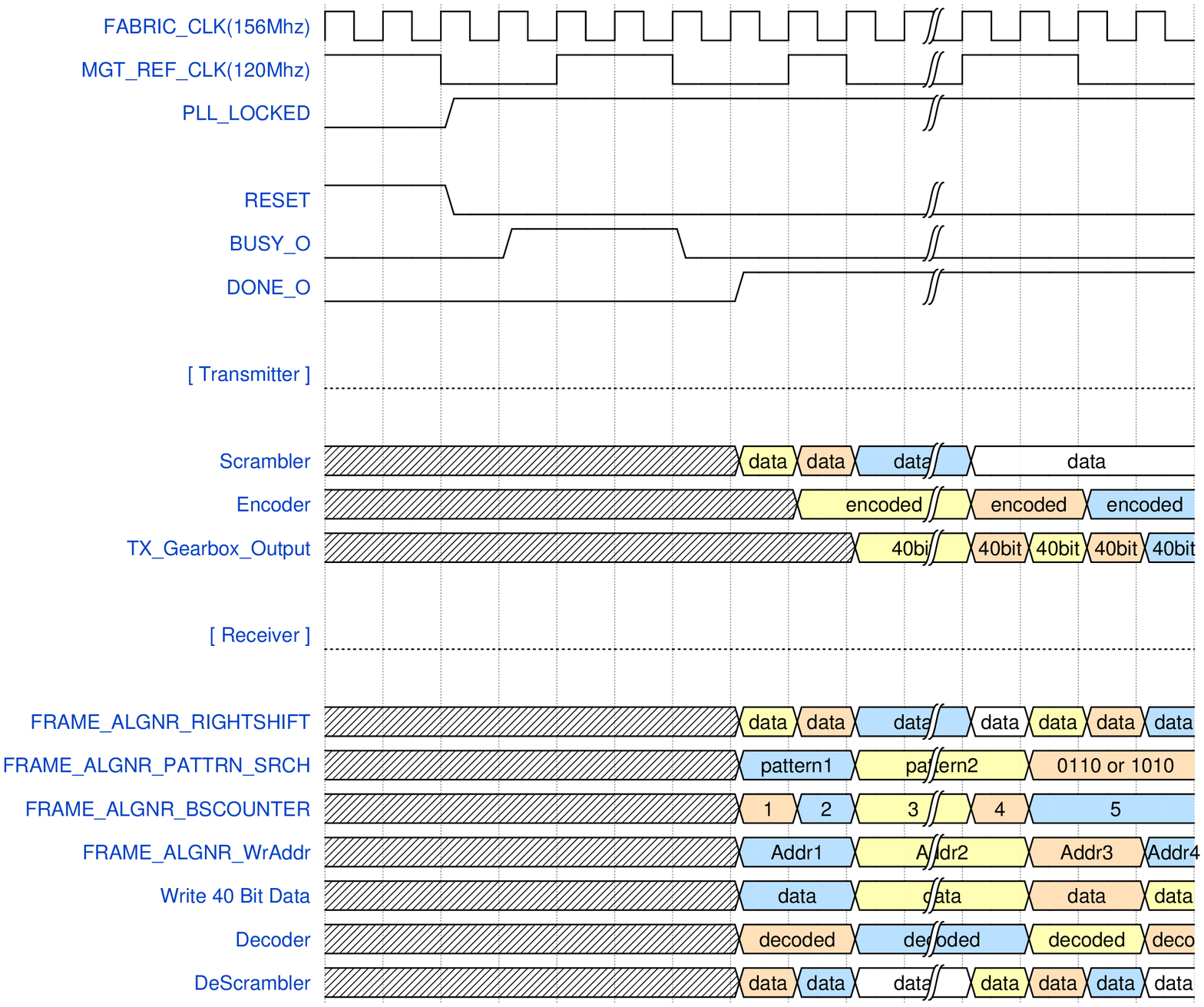}
\caption{Timing diagram of the transmitter and receiver signals }
\vspace*{-8pt}
\label{fig:GBTTimingTx}
\end{figure*}

The full prototype of DAQ chain is implemented in the Xilinx Kintex-7 boards (KC705 from Avnet) using the Xilinx ISE 14.5 platform and VHDL language. We have used an external jitter cleaned clock source (CDCE62005EVM of TI) to drive MGT of two Kintex-7 boards. The timing diagram of the transmitter and the receiver side are given in Figure~\ref{fig:GBTTimingTx}. Name of the signals and their  function are given in Table~\ref{table:SignalName}.

\begin{table*}[htb]
\centering

\scalebox{0.82}{
\begin{tabular}  {|c|c|c|} 
\hline
Signal Name & Function & Use \\ 
\hline
Fabric\_Clk & Use to drive different blocks in DAQ & Use both in Transmission and Receiver side \\ 
\hline
MGTREF\_Clk & Use to drive MGT & Use both in Transmission and Receiver side \\ 
\hline
PLL\_Clk & Use to drive MGT & Use both in Transmission and Receiver side \\ 
\hline
PLL\_Locked & Output of PLL. It indicates PLL generate stable clock & Use both in Transmission and Receiver side\\
\hline
RESET & Use to reset the whole system & Use both in transmission and Receiver side\\
\hline
BUSY\_O & Becomes high when System enters a process before ready & Use both in transmission and Receiver side\\
\hline
Scrambler & This signal contains the data of output of scrambler block & Use in Transmission side only\\
\hline
Encoder & This signal contains the data after BCH encoding & Use in Transmission side only\\
\hline
MUX\_Output & This signal contains output of MUX block which is 40 bit width & Use in Transmission side only\\
\hline
FRAME\_ALIGNR\_RIGHTSHIFT & store the receive data after shifting right by one bit & Use in Receiver side only\\
\hline
FRAME\_ALIGNR\_PATTERN\_SEARCH & Check whether header is matched or not & Use in Receiver side only\\
\hline 
FRAME\_ALIGNR\_BSCOUNTER & Store the output of counter until header is not matched & Use in Receiver side only\\
\hline 
Header\_LOCK\_O & Becomes high when the frame is locked & Use in Receiver side only\\
\hline
FRAME\_ALIGNR\_WrAddr & store the address of RAM where receive data will be written & Use in Receiver side only\\
\hline
RAM\_ENABLE & Becomes high when RAM is Ready to perform & Use in Receiver side only\\
\hline
Write 40 bit Data & Store 40 bit data which is to be written in RAM &  Use in Receiver side only\\
\hline
DECODER\_ENABLE & Becomes high when BCH decoder is ready to perform & Use in Receiver side only\\
\hline 
DECODER & Contains the decoded data & Use in Receiver side only\\
\hline
DESCRAMBLER & Contains output data of descrambler block & Use in Receiver side only\\  
\hline
\end{tabular}}
\caption{Description of the signals used in timing diagram}
\label{table:SignalName}
\end{table*}


%


\section{Experimental Setup and performance} \label{PerformenceEvoluation}
The block diagram and experimental setup of the system are shown Figure~\ref{fig:PCIeSetup}. We achieved maximum bit rate 4.8 Gbps in our system. In standard mode, a frame contains only 52 bits of data, 64 bits for error correction (FEC) and 4 bits of header. 64 bits for FEC can correct up to 16 bits of error, as it is applied on 8 encoder blocks in parallel (2 bit error correction for each block). In extra wide bus frame format, where error correction code is not used, so the frame can carry $(52+64=116)$ bits of data, out of 120 bits frame.
So the data rate achieved considering only the data field (D) in standard mode is: \\
$ 40 MHz \times 52 bits = 2.08 Gbps$ \\
Similarly, in non-error correctional mode (extra wide bus mode), data rate is measured: \\
$40 MHz \times 116 bits = 4.64 Gbps$ \\
So, the data transfer efficiency for the above mention two modes are $(2.08/4.80)\times 100 = 43.33\%$ and $(4.64/4.80 = 96.6\% $ respectively.

Resource utilization for each functional block of the system including critical time delay is given in Table~\ref{table:resource}. 
\begin{table*}[htb]
\centering

\vspace{10 pt}
\scalebox{1.0}{
\begin{tabular}{|c|c|c|c|c|c|c|}
\hline 
Board & Module Name & Slice Register & Slice LUTs & LUT-Flip Flop & BRAM & Critical path (ns)\\ 
\hline 
\multirow{10}{*}{\rotatebox{0}{Kintex 7-325t}} & BCH Encoder(15,7,2)  & 7//407600 & 951/203800& 0 &7/951 &0.373\\ \cline{2-7}
 & BCH Decoder(15,7,2) & 135/407600 & 446/203800 & 0 & 119/462(25\%) & 0.985\\
 \cline{2-7}
 & Scrambler & 52 & 53 & 5 & 0 & 0.786 \\
 \cline{2-7}
  &Descrambler &104 & 56 & 5 & 0 & 0.689 \\
 \cline{2-7}
&Interleaver &44 & 40 & 40 & 0 & 0.905\\
 \cline{2-7}
&DeInterleaver &201 & 82 & 80 & 0 & 0.634 \\
 \cline{2-7}
&Frame Aligner &115& 308 & 72 & 0 & 2.294 \\
 \cline{2-7}
 &PCIe &5882 & 5287 & 2694 & 10 & 3.875 \\
 \cline{2-7}
 &Top Module&3665 &9003 & 1998 & 5 & 8.62 \\
  & Without PCIe & & &  & & \\
 \cline{2-7}
 &Top Module&8360 & 8555 & 3779 & 26 & 11.455 \\
 & With PCIe & & & & & \\
 \cline{2-7}  
 \hline
\end{tabular} }
\caption{Resource Utilization}
\label{table:resource}
\end{table*}  

Critical time is the maximum delay time, to get the output of a circuit block after the input is given. Power consumption is calculated using Xilinx Xpower tool and is given in  Table~\ref{table:powerconsumtion}. 
\begin{table}[htb]
\centering

\scalebox{0.7}{
\begin{tabular}{|c|c|c|c|}
\hline 
Board & Module Name & Logic Power(mW) & Signal Power(mW)\\ 
\hline 
\multirow{10}{*}{\rotatebox{0}{Kintex 7-325t}} & BCH Encoder(15,7,2)  & 0.02 & 0.01\\ \cline{2-4}
 & BCH Decoder(15,7,2) & 0.05 & 0.07 \\
 \cline{2-4}
 & Scrambler & 0.04 & 0.00 \\
 \cline{2-4}
  &Descrambler &0.01 & 0.00 \\
 \cline{2-4}
&Interleaver &0.01 & 0.01\\
 \cline{2-4}
&DeInterleaver & 0.01& 0.02 \\
 \cline{2-4}
&Frame Aligner &1.34& 1.07 \\
 \cline{2-4}
 &PCIe &253.24 & 45.55 \\
 \cline{2-4}
 &Top Module& 474.18 & 2.91 \\
  & Without PCIe & &  \\
 \cline{2-4}
 &Top Module& 304.24 & 56.31 \\
 & With PCIe & &  \\
 \cline{2-4}  
 \hline
\end{tabular} }
\caption{Module wise power consumption}
\label{table:powerconsumtion}
\end{table}   

The video link of the real lab setup is given here. \url{https://vimeo.com/113255103}
\section{Conclusion}
In this work we have proposed a novel DAQ design for HEP experiments. The proposed DAQ supports high speed (Gbps) optical data communication and also achieves multi-bit error correction. The DAQ design has been implemented on Xilinx Kintex-7 board and real test setup has been developed involving board to board communication and PCIe interfacing with a host PC. A detailed performance analysis of the DAQ implementation is presented in terms of timing diagram, resource utilization and critical path delay for of each blocks (FPGA) and power consumption. The proposed DAQ design and its implementation involving optical data communication and multi-bit error correction capability can be considered as first of its kind and can serve as a benchmark design in HEP DAQ.
\bibliographystyle{abbrv}
\bibliography{IEEEexample}

\begin{thebibliography}{1}

\bibitem{Bose196068}
R.~Bose and D.~Ray-Chaudhuri.
\newblock On a class of error correcting binary group codes.
\newblock {\em Information and Control}, 3(1):68 -- 79, 1960.

\bibitem{chien:search}
R.~Chien.
\newblock Cyclic decoding procedures for bose- chaudhuri-hocquenghem codes.
\newblock {\em Information Theory, IEEE Transactions on}, 10(4):357--363, 1964.

\bibitem{Nashiyama:SEU}
I.~Nashiyama, T.~Nishijima, H.~Sekiguchi, Y.~Shimano, and T.~Goka.
\newblock Study of basic mechanisms of single event upset using high-energy
  microbeams.
\newblock {\em Nuclear Instruments and Methods in Physics Research Section B:
  Beam Interactions with Materials and Atoms}, 54(1–3):407 -- 410, 1991.

\bibitem{Xie:Tripple:BCH}
X.~Zhi-yuan, L.~Na, and L.~Le-le.
\newblock New decoder for triple-error-correcting binary bch codes.
\newblock In {\em Industrial Electronics and Applications, 2008. ICIEA 2008.
  3rd IEEE Conference on}, pages 1426--1429, June 2008.

\end{thebibliography}
\end{document}